\documentclass[10pt, paper=a4, UKenglish]{article}
\usepackage{graphicx}
\usepackage{amsmath}
\usepackage{multirow}
\usepackage{stmaryrd}
\usepackage{placeins}

\def\Title#1{\begin{center} {\Large #1 } \end{center}}
\def\Author#1{\begin{center}{ \sc #1} \end{center}}
\def\Address#1{\begin{center}{ \it #1} \end{center}}

\newcommand\pubblock{\rightline{\begin{tabular}{l} Proceedings of the CTD 2023\\ \pubnumber\\
         \pubdate  \end{tabular}}}

\newenvironment{Abstract}{\begin{quotation} \begin{center} 
             \large ABSTRACT \end{center}\bigskip 
      \begin{large}}{\end{large} \end{quotation}}

\newenvironment{Presented}{\begin{quotation} \begin{center} 
             PRESENTED AT\end{center}\bigskip 
      \begin{center}\begin{large}}{\end{large}\end{center} \end{quotation}}

\def\Acknowledgements{\bigskip  \bigskip \begin{center} \begin{large}
      \bf ACKNOWLEDGEMENTS \end{large}\end{center}}

\def\beq{\begin{equation}}
\def\eeq#1{\label{#1}\end{equation}}
\def\eeqn{\end{equation}}

\def\beqa{\begin{eqnarray}}
\def\eeqa#1{\label{#1}\end{eqnarray}}
\def\eeqan{\end{eqnarray}}

\let\bar=\overbar

\def\Dslash{\not{\hbox{\kern-4pt $D$}}}
\def\dslash{\not{\hbox{\kern-2pt $\del$}}}

\def\msb{{\bar{\ssstyle M \kern -1pt S}}}

\usepackage{color}
\usepackage{lineno}
\usepackage{subfig}
\usepackage{hyperref}

\newcommand\pubnumber{PROC-CTD2023-34}

\newcommand\pubdate{\today}

\def\affiliation{
On behalf of the LHCb real-time analysis project}

\newcommand{\conference}{Connecting the Dots Workshop (CTD 2023)\\
October 10-13, 2023}

\usepackage{fancyhdr}
\definecolor{mygrey}{RGB}{105,105,105}
\begin{document}


\large
\begin{titlepage}
\pubblock

\vfill
\Title{Graph Neural Network-Based Pipeline for Track Finding in the Velo at~LHCb}
\vfill

\Author{Anthony~Correia$^1$, Fotis~I.~Giasemis$^{1, 2}$, Nabil~Garroum$^1$, Vladimir~Vava~Gligorov$^1$, Bertrand Granado$^2$}

\Address{$^1$ LPNHE, Sorbonne Université, CNRS/IN2P3, Paris, France}
\Address{$^2$ LIP6, Sorbonne Université, CNRS, Paris, France}
\vfill
\Address{\affiliation}
\vfill

\begin{Abstract}
Over the next decade, increases in instantaneous luminosity and detector granularity will amplify the amount of data that has to be analysed by high-energy physics experiments, whether in real time or offline, by an order of magnitude.
The reconstruction of charged particle tracks, which has always been a crucial element of offline data processing pipelines, must increasingly be deployed from the very first stages of the real time processing to enable experiments to achieve their physics goals.
Graph Neural Networks (GNNs) have received a great deal of attention in the community because their computational complexity scales nearly linearly with the number of hits in the detector, unlike conventional algorithms which often scale quadratically or worse.
This paper presents \texttt{ETX4VELO}, a GNN-based track-finding pipeline tailored for the Run 3 LHCb experiment's Vertex Locator, in the context of LHCb's fully GPU-based first-level trigger system, Allen. 
Currently implemented in Python, \texttt{ETX4VELO} offers the ability to reconstruct tracks with shared hits using a novel triplet-based method. When benchmarked against the traditional track-finding algorithm in Allen, this GNN-based approach not only matches but occasionally surpasses its physics performance. In particular, the fraction of fake tracks is reduced from over 2\% to below 1\% and the efficiency to reconstruct electrons is improved.
While achieving comparable physics performance is a milestone, the immediate priority remains implementing \texttt{ETX4VELO} in Allen in order to determine and optimise its throughput, to meet the demands of this high-rate environment.
\end{Abstract}

\vfill

\begin{Presented}
\conference
\end{Presented}
\vfill
\end{titlepage}
\def\thefootnote{\fnsymbol{footnote}}
\setcounter{footnote}{0}
%

\normalsize 


\section{Introduction}
\label{intro}

The LHCb Run~3 detector~\cite{LHCb:2012doh} is a multi-purpose particle physics detector at the Large Hadron Collider (LHC). Proton bunches collide every 25 ns within its Vertex Locator (Velo). The resulting particles are detected in a forward region near the beamline, specifically within the acceptance region defined by $\eta \in \left[2, 5 \right]$, where $\eta$ represents the pseudo-rapidity.

\begin{figure}[!htb]
  \centering
  \includegraphics[width=0.70\linewidth]{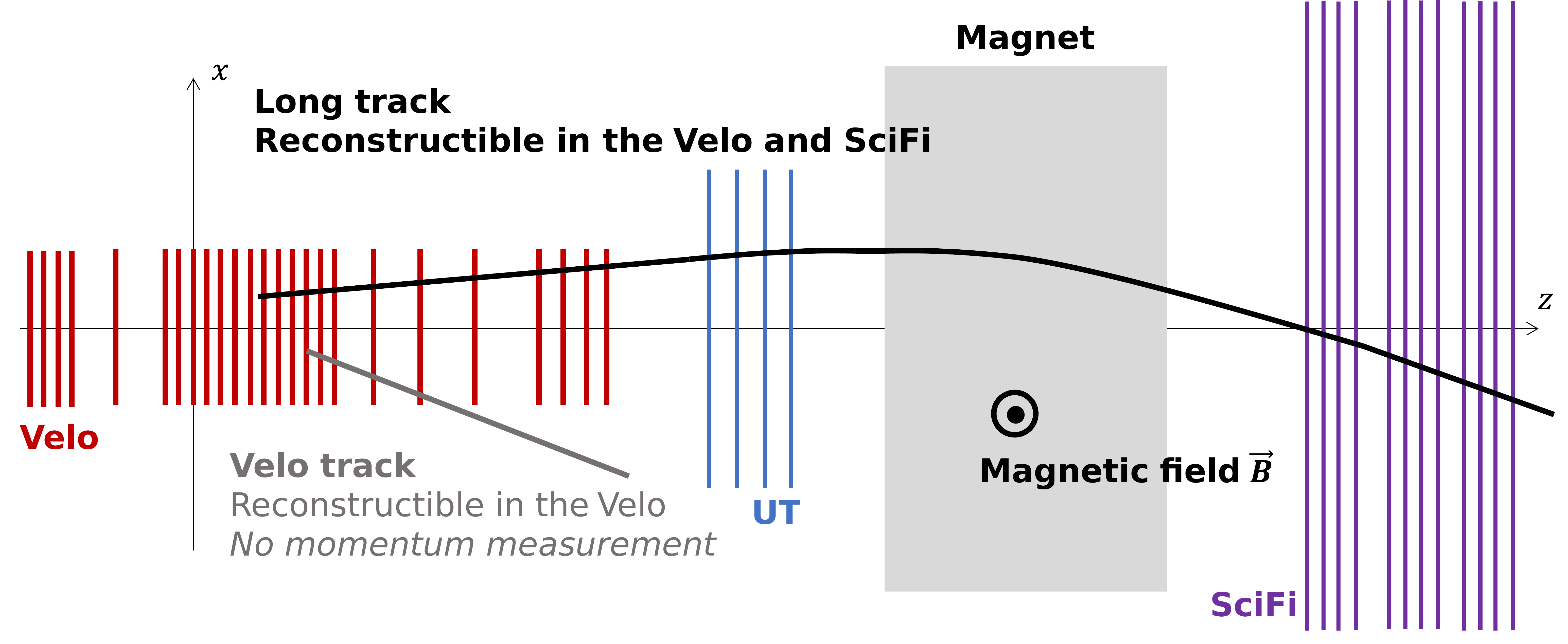}
  \caption{Sketch of the LHCb tracking system in Run~3.}
  \label{fig:tracking_system}
\end{figure}

Central to the LHCb is its tracking system, illustrated in Figure~\ref{fig:tracking_system}. This system consists of:
\begin{itemize}
    \itemsep0em
    \item The \textbf{Vertex Locator (Velo)}~\cite{Bediaga:2013tje}, surrounding the proton-proton interaction region, encompasses $n_{\text{planes}} = 26$ planes of $55\times55$ ${\mu m}^2$ silicon pixels.
    \item The \textbf{Upstream Tracker (UT)}~\cite{LHCb:2014uqj}, placed before the magnet station, includes 4 planes of silicon strips.
    \item The \textbf{Scintillating Fibre Tracker (SciFi)}~\cite{LHCb:2014uqj}, located after the magnet station, is composed of 12 planes of scintillating fibres.
\end{itemize}

The LHCb subdetectors collectively produce a data rate of up to 5 TB/s. To curtail this before storage during Run~3, LHCb deployed Allen, a GPU-based online trigger~\cite{LHCb:2020kay}, which trims the rate to a more manageable 70-200 GB/s. Using around 340 GPUs, Allen partially reconstructs events, achieving better selection efficiency than the FPGA-based hardware trigger used by LHCb during Run~1 and Run~2. The data is then buffered for real-time alignment and calibration before a second, x86-based trigger performs a full event reconstruction and selects events containing physics signals of interest for analysis. This refined data is then stored at a rate of 10 GB/s.

In this environment, Allen and LHCb offers a pragmatic platform for developing Neural Network-based algorithms on GPUs, with a specific emphasis on high-throughput. Using Allen's existing GPU algorithms as a reference, the track-finding algorithms in development can be refined and compared both in terms of physics and throughput performance.

This paper introduces \texttt{ETX4VELO}, a Graph Neural Network (GNN)-based track-finding pipeline for the Velo. It is based on the Exa.TrkX collaboration's approach~\cite{ExaTrkX:2021abe}, which was originally tailored for $4\pi$ tracking detectors in a magnetic field, akin to ATLAS and CMS, in the context of the High-Luminosity upgrade of the LHC (HL-LHC). Crucially, the Exa.TrkX pipeline exhibited a near-linear relationship between throughput and hit count, contrasting with the quadratic nature of conventional algorithms. This is of particular interest in the context of LHCb's upcoming upgrades, which will enhance instantaneous luminosity and detector granularity.

\section{Track Topologies in the Velo}
\label{track_topology}

In simulation, there are, on average, 150 particles in the Velo acceptance to reconstruct, and 2,200 hits per event. Around 15\% of the hits are spilled over from prior events and are therefore noise.

The Velo tracks exhibit the following key characteristics:
\begin{itemize}
    \itemsep0em
    \item \textbf{Linear Trajectories:} The tracks are straight lines as there is no magnetic field within the Velo.
    \item \textbf{Reconstructibility Criterion:} A track is deemed \textit{reconstructible in the Velo} if it leaves hits in at least three Velo layers~\cite{Li:2021oga}.
    \item \textbf{Forward Orientation:} The tracks to be reconstructed are forward-oriented, with $\eta \in \left[2, 5\right]$, for interaction with subsequent LHCb subdetectors. Although backward-oriented tracks (with negative $\eta$) are also relevant, particularly for primary vertex reconstruction, they are not evaluated in this paper.
    \item \textbf{Skipped Plane:} In simulations, a particle may skip a plane in 1\% of cases, attributed to the inefficiency of the detector plane.
    \item \textbf{Multiple Hits per Plane:} A track can generate more than one hit per plane.\footnote{A Velo plane consists of four overlapping sensor layers, displaced in $z$, collaboratively covering the desired acceptance in the $\left(x, y\right)$ plane. Multiple hits on a plane primarily arise from different overlapping sensor layers within a plane.}
    \item \textbf{Shared Hits:} Tracks may intersect, leading to a \textit{shared hit}.
    \item \textbf{Positron-Electron Pairs:} Material interactions frequently produce positron-electron pairs, resulting in two tracks that initially share hits before diverging.
\end{itemize}

\begin{figure}[!htb]
  \centering
  \includegraphics[width=0.25\linewidth]{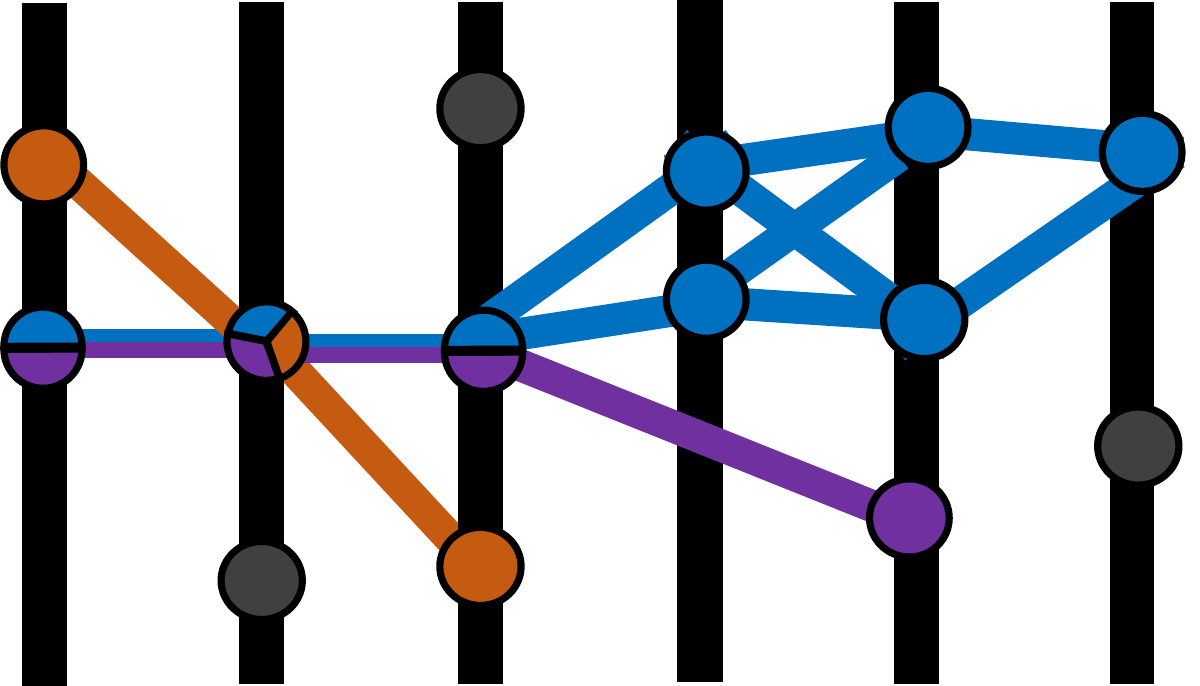}
  \caption{Simplified example of tracks to be reconstructed in the Velo. The blue and purple tracks share three hits prior to diverging. The purple track jumps from plane 3 to 5, missing 4. The orange track intersects the blue and purple tracks. Dark points represent hits unassociated with any particle. When considering the hits as graph nodes, lines between hit nodes represent the genuine edges, as defined in this work.}
  \label{fig:tracks}
\end{figure}

Given these characteristics, an effective track-finding algorithm for the Velo should be capable of reconstructing the particle tracks illustrated in Figure~\ref{fig:tracks}.

\section{ETX4VELO Pipeline}
\label{pipeline}

\begin{figure}[!htb]
  \centering
  \includegraphics[width=0.98\linewidth]{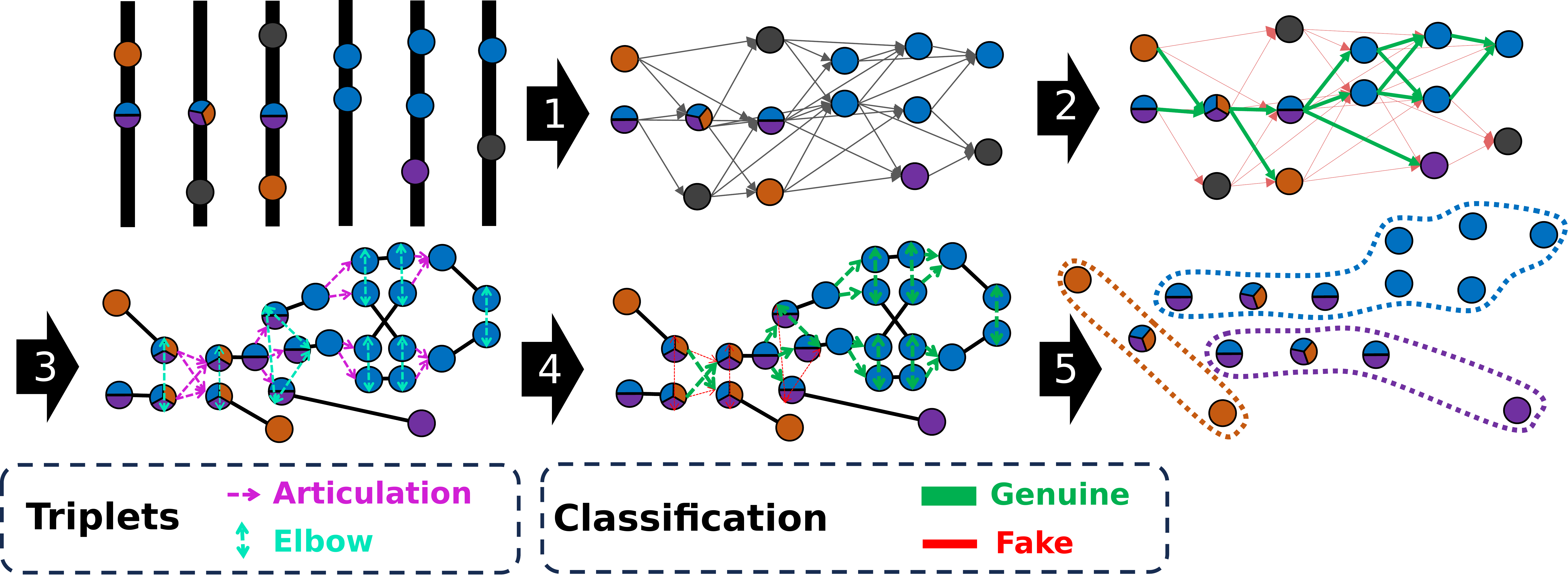}
  \caption{Illustration of the pipeline's 5 stages, beginning with hits from the minimalist example in Figure~\ref{fig:tracks}. Steps entail (1) building a rough hit graph, (2) classifying its edges and discarding fakes (in red), (3) constructing the edge graph with edge-to-edge connections called triplets, (4) classifying and removing fake triplets, and finally, producing the tracks.}
  \label{fig:pipeline}
\end{figure}

This section outlines \texttt{ETX4VELO}, a five-step track-finding pipeline, as illustrated in Figure~\ref{fig:pipeline}.

\begin{enumerate}
    \item \textbf{Hit Graph Construction}: A preliminary directed graph of connected hits $G^{\text{hit}}_{\text{rough}}$ is built. The target edges are connections between hits from the same particle on adjacent planes. The hit coordinates are embedded using a Dense Neural Network (DNN) and the edges are built by local $k$-Nearest Neighbours ($k$-NN) applications. Further details are in Section~\ref{pipeline:graph_building}.
    \item \textbf{Edge Classification}: A GNN scores the edges in $G^{\text{hit}}_{\text{rough}}$ between 0 (fake) and 1 (genuine). Those below $s_{\text{edge, min}}$ are discarded, forming the purified hit graph $G^{\text{hit}}_{\text{purified}}$. Further details are in Section~\ref{pipeline:gnn_edge}.
    \item \textbf{Edge Graph Construction}: 
    An edge graph, $G^{\text{edge}}$, is derived from $G^{\text{hit}}_{\text{purified}}$. Here, nodes of $G^{\text{edge}}$ are edges from $G^{\text{hit}}_{\text{purified}}$ and two edges connect if they share a hit. The edge-edge connections, are called \textit{triplets} since they involve three hits. Further details are in Section~\ref{pipeline:edge_graph_construction}.
    \item \textbf{Triplet Classification}: Triplets within $G^{\text{edge}}$ undergo classification. Those scoring below $s_{\text{triplet, min}}$ are removed, producing the purified edge graph $G^{\text{edge}}_{\text{purified}}$. Further details are in Section~\ref{pipeline:gnn_triplet}.
    \item \textbf{Track Construction}: The tracks are reconstructed from the purified edge graph  $G^{\text{edge}}_{\text{purified}}$ through an algorithm involving a Weakly Connected Component (WCC) stage, as detailed in Section~\ref{pipeline:track_building}.
\end{enumerate}

Compared to the Exa.TrkX pipeline, which directly follows its first two stages with a WCC algorithm, \texttt{ETX4VELO} introduces three additional stages to handle tracks with shared hits.


\subsection{Rough Graph Construction}
\label{pipeline:graph_building}

To limit the graph size, edges created during the graph construction stage are permitted to skip at most one plane. Edges are also directed towards increasing plane number (and $z$-coordinate).

Initially, the hit coordinates $\left(r, \phi, z, \text{plane number}\right)$\footnote{Later analyses showed omitting the plane number from DNN input did not affect performance.} are normalised and input into a DNN. This DNN embeds hits into a $n_{\text{dim}}$-dimensional space, set as $n_{\text{dim}} = 4$, designed to bring connected hit pairs closer while distancing disconnected ones. The embedding DNN comprises 3 hidden layers of 128 hidden units with hyperbolic tangent activation, each followed by a normalisation layer, and an output layer without activation. In total, the network has approximately 35,000 parameters.

The rough graph $G^{\text{hit}}_{\text{rough}}$ is built by identifying the $k_{\text{max}}$-nearest neighbours (in the embedding space) of every hit in plane $p \in \left\llbracket 1, n_\text{planes} - 1 \right\rrbracket$, constrained within the subsequent two planes $p + 1$ and $p + 2$, and a maximum squared distance of $d^2_{\text{max}}$. Practically, a $k_{\text{max}}$-Nearest Neighbour ($k_{\text{max}}$-NN) algorithm implemented in \texttt{faiss}~\cite{Johnson:2021tbd} determines hits' neighbours between plane $p$ and planes $p+1$ and $p+2$. Only neighbours within the $d^2_{\text{max}}$ boundary are retained. The values of $k_{\text{max}}$ and $d^2_{\text{max}}$ are optimisable parameters.
\newline

For training the DNN, a dataset consisting of (1) random hit pairs, (2) all edge-connected hit pairs, and (3) challenging fakes identified by the 25 local $k_\text{max}$-NN procedure with $k_\text{max} = 50$ and $d^2_\text{max} = 0.015$, is utilised. 
This results in a dataset of $n_{\text{genuine}}$ connected hits and $n_{\text{fake}}$ disconnected ones.
The distances $\left\{d_{\text{genuine},i}, \forall i \in \llbracket 1, n_{\text{genuine}}\rrbracket\right\}$ and $\left\{d_{\text{fake},j}, \forall j \in \llbracket 1, n_{\text{fake}}\rrbracket\right\}$  between the embedded genuine and fake hit pairs, respectively, are then computed.

The DNN training minimises a pairwise hinge embedding loss defined as 
\begin{equation}
    \mathcal{L} =  \mathcal{L}_{\text{fake}} + w_\text{genuine} \times \mathcal{L}_{\text{genuine}}\,,
\end{equation}
with a weight $w_\text{genuine} = 3$ to emphasise genuine edge inclusion over fake edge exclusion. The optimization of hyperparameters such as $m$ and $w_\text{genuine}$ will be pursued in future work.

The objective of the following loss equations is to reduce the distance between connected hits and increase the distance (up to a threshold $m$) between disconnected ones:
\begin{equation}
    \mathcal{L_{\text{genuine}}} = \frac{1}{n_{\text{genuine}}} \sum_{i = 1}^{n_\text{genuine}} d^2_{\text{genuine},i}
     \quad\text{and}\quad 
    \mathcal{L_{\text{fake}}} = \frac{1}{n_{\text{fake}}} \sum_{j = 1}^{n_\text{fake}} \max{\left(0, m - d^2_{\text{fake},j}\right)}\,,
\end{equation}
with the margin $m$ fixed at $0.010$.

\subsection{Edge Classification}
\label{pipeline:gnn_edge}

\begin{figure}[!htb]
  \centering
  \includegraphics[width=\linewidth]{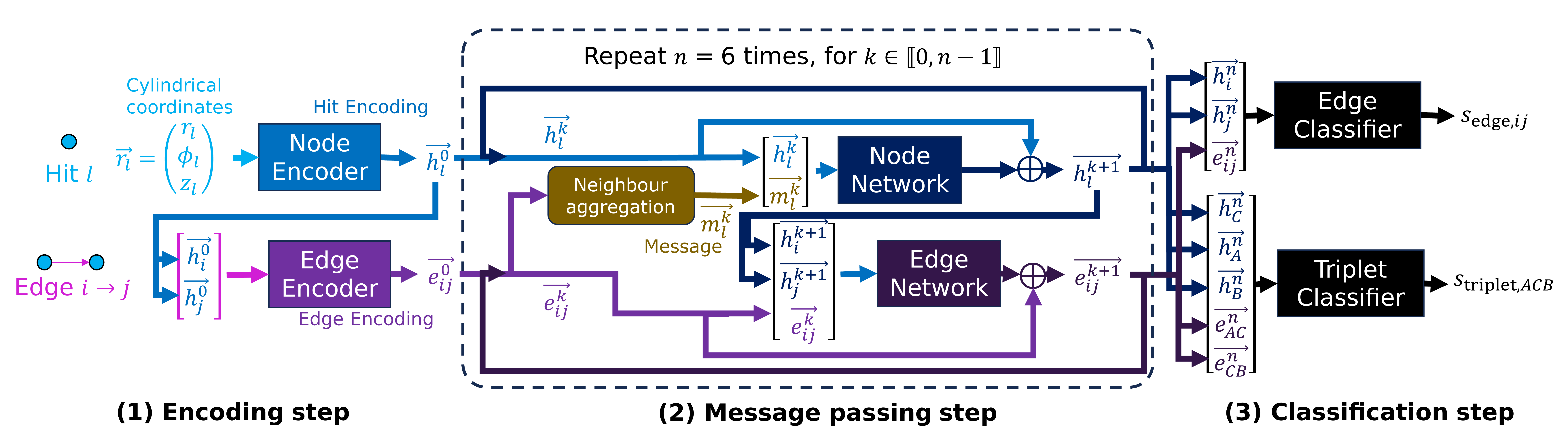}
  \caption{Schematic of the GNN architecture, highlighting: (1) hit and edge encodings, (2) iterative message passing refinement, and (3) subsequent edge and triplet classifications. The node and edge encoders and networks and the edge and triplet classifiers are DNNs.}
  \label{fig:gnn}
\end{figure}

The edges are classified using a GNN that incorporates a 6-step message passing mechanism. Notably, the message-passing aggregation scheme has been updated from the original Exa.TrkX pipeline. The architecture of this GNN is illustrated in Figure~\ref{fig:gnn}.

Each hit, denoted as $l$, is encoded from its normalised cylindrical coordinates $r_l = \left(r_l, \theta_l, z_l\right)$ through a 3-layer DNN (\textit{hit encoder}) into a 256-dimensional representation, denoted as $h_l^{0}$.

For every edge connecting hits $i$ and $j$, its 256-dimensional encoding $e_{ij}^0$ is formed by concatenating the associated hit encodings $h^0_i$ and $h^0_j$, which is then processed through a 3-layer DNN (\textit{edge encoder}).

During the $n$-step message passing phase, with $n = 6$, at each step $k\in\left\llbracket 0, n - 1\right\rrbracket$, a \textit{message} vector $m^{k}_{l}$ is derived for every hit $l$.
The message vector aggregates encodings from both its preceding (left, $q_L \to l$) and succeeding (right, $l \to q_R$) edges. Specifically:
\begin{equation}
    m^{k}_{l} = \left[
    \text{max}\left(\left\{e^{k}_{q_{L}l}\right\}_l\right)
    ,\,
    \text{sum}\left(\left\{e^{k}_{q_{L}l}\right\}_l\right)
    ,\,
    \text{max}\left(\left\{e^{k}_{lq_{R}}\right\}_l\right)
    ,\,
    \text{sum}\left(\left\{e^{k}_{lq_{R}}\right\}_l\right)
    \right]\,.
\end{equation}
This distinct aggregation for preceding and succeeding planes enhances performance.

After message derivation, the hit encodings $h^{k}_{l}$ upgrade to $h^{k + 1}_{l}$ using a 3-layer DNN (\textit{node network}). This network features a residual connection and ingests the current hit encoding $h^{k}_{l}$ and the message $m^{k}_{l}$. Similarly, edge encodings $e^k_{ij}$ are refined using a 6-layer DNN (\textit{edge network}) with a residual connection, from the refined hit encodings $h^{k + 1}_{i}$ and $h^{k + 1}_{j}$ alongside the current edge encoding $e^k_{ij}$.

After message passing, each edge between hits $i$ and $j$ is scored between 0 (indicative of fakeness) and 1 (indicative of genuineness) by a 3-layer DNN (\textit{edge classifier}) using the refined hit encodings $h^{n}_i$ and $h^{n}_j$ and edge encodings $e^{n}_{ij}$. Only those surpassing the score threshold, $s_{\text{edge, min}}$, persist in the purified hit graph $G^{\text{hit}}_{\text{purified}}$.

All the DNNs have 256 hidden units per layer and use the SiLU activation~\cite{elfwing_sigmoid-weighted_2018}, each followed by a normalisation layer. The output layers of the classifiers deploy a sigmoid activation.

The GNN training encompasses the triplet classifier, which is detailed further in Section~\ref{pipeline:gnn_triplet}. In total, it has approximately two million parameters.

\subsection{Edge Graph Construction}
\label{pipeline:edge_graph_construction}

Applying a WCC algorithm directly to the hit graph results in tracks that share common hits being merged. This is particularly detrimental for positron-electron pairs that share their first hits before diverging, as detailed in Section~\ref{track_topology}. This leads to an unsatisfactory electron track-finding performance.

The hit graph, by its construction, only encodes hit-to-hit connections and lacks the capability to disentangle tracks that share hits. Transitioning to connections between the edges themselves offers a solution. Thus, we introduce the \textit{edge graph} $G^{\text{edge}}$, where nodes correspond to the edges from the hit graph, and connect if their corresponding edges share a common hit.

\begin{figure}[!htb]
  \begin{center}
  \subfloat[]{\includegraphics[width=0.16\linewidth]{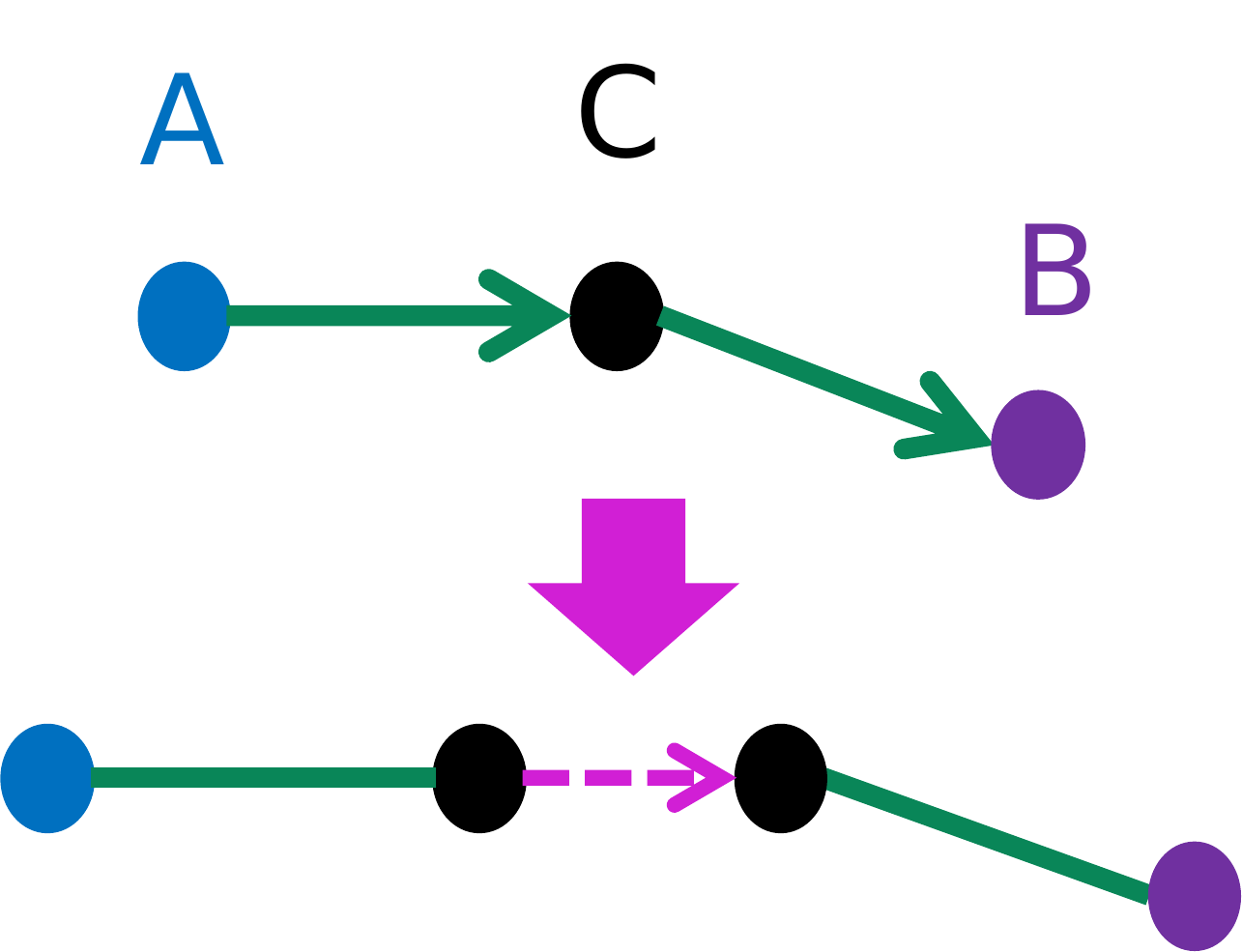}}
  \qquad
  \subfloat[]{\includegraphics[width=0.15\linewidth]{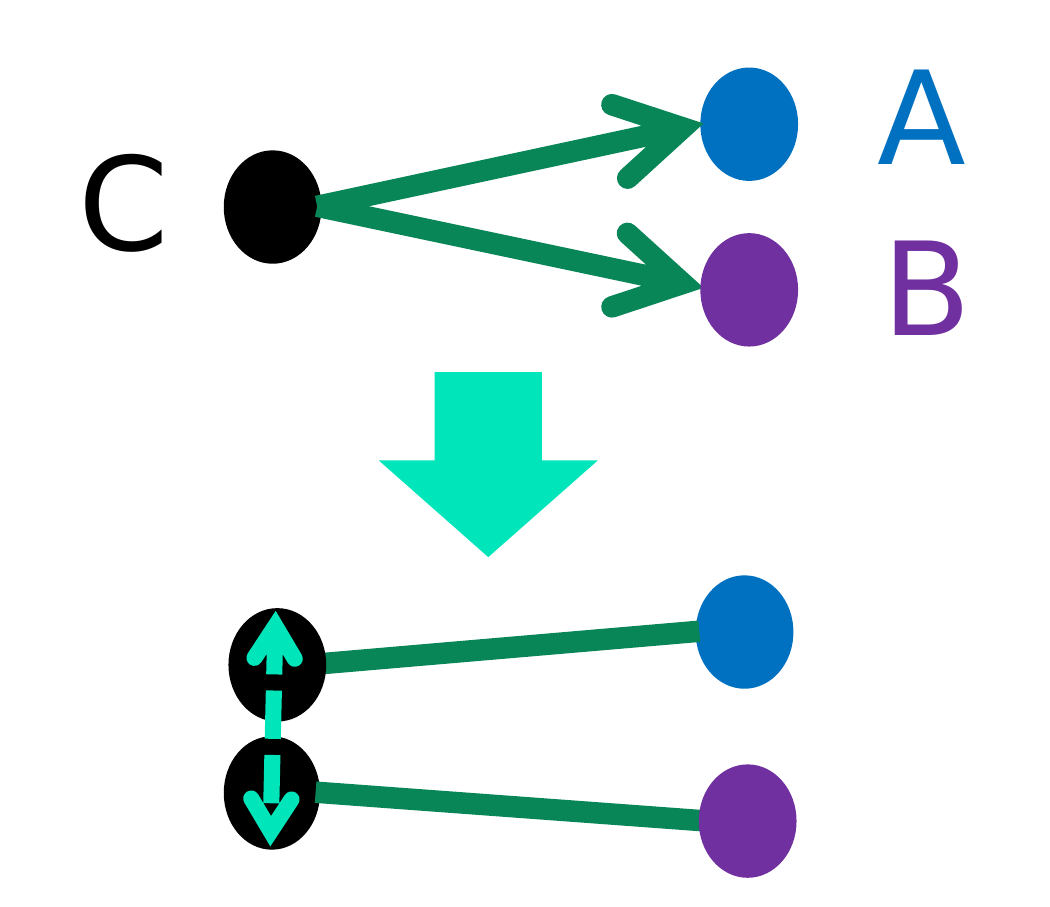}}
  \qquad
  \subfloat[]{\includegraphics[width=0.15\linewidth]{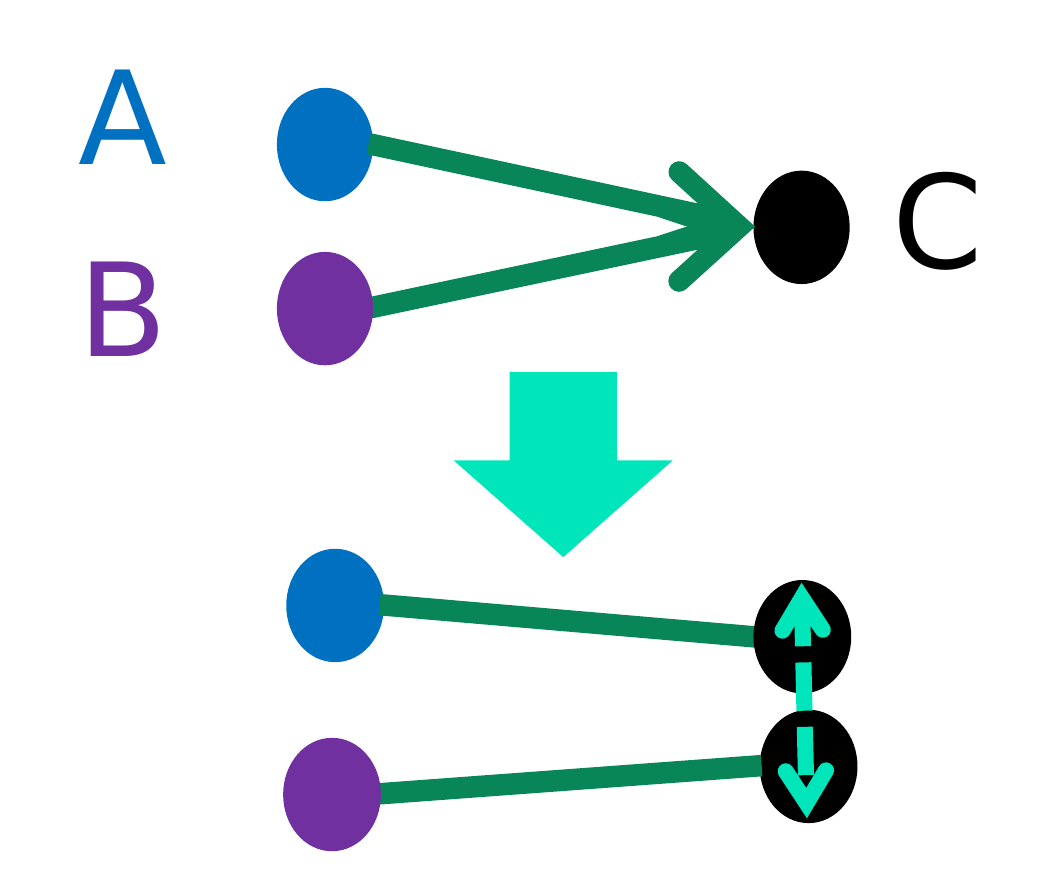}}
  \caption{Visual representation of the three triplet configurations in the edge graph: (a) the articulation, (b) the left elbow and (c) the right elbow.}
  \label{fig:triplets}
  \end{center}
\end{figure}

An edge-edge connection in this graph represents a set of two edges connected by a shared hit $C$, and is therefore called \textit{triplets}~\cite{ExaTrkX:2020apx}. Considering the directionality of these edges (towards increasing plane numbers), the triplets in the edge graph can be categorised into three types, visually depicted in Figure~\ref{fig:triplets}:
\begin{itemize}
    \itemsep0em
    \item \textbf{Articulation}: Two sequential edges $A \to C$ and $C \to B$, maintaining the order ${\text{plane}_A < \text{plane}_C < \text{plane}_B}$.
    \item \textbf{Left Elbow}: Two edges $C\to A$ and $C\to B$ emanating from a common left hit $C$.
    \item \textbf{Right Elbow}: Two edges  $A\to C$ and $B \to C$ converging at a common right hit $C$.
\end{itemize}

For scenarios with tracks sharing a hit $S$, not all triplets with the common hit $C = S$ will be genuine. Such ambiguous triplets necessitate classification and potential exclusion, as elaborated in the next section.

\subsection{Triplet Classification}
\label{pipeline:gnn_triplet}

To classify the triplets in the edge graph $G^{\text{edge}}$, the node and edge encodings from the GNN-based edge classifier described in Section~\ref{pipeline:gnn_edge} are utilised instead of invoking another GNN inference.

For a triplet defined by hits $(C, A, B)$, where $C$ is the common hit, the encodings of these hits and of the edges $A\leftrightarrow C$ and $B \leftrightarrow C$ are input to a deep neural network (DNN), termed the \textit{triplet classifier}. This DNN outputs a score between 0 (fake) and 1 (genuine). For left and right elbows, which lack a strict hit order between $A$ and $B$, the triplet score is the average of the two scores for both possible orderings.
Triplets are then filtered based on a threshold score $s_\text{triplet, min}$ to exclude the non-genuine ones.

The overall GNN aims at both edge and triplet classification, by minimising the combined loss: $\mathcal{L} = \mathcal{L}_{\text{edges}} + \mathcal{L}_{\text{triplets}}$. During training, edges with scores below 0.5 are discarded prior to triplet construction, concentrating on challenging triplet cases. Both edge and triplet losses employ the sigmoid focal loss~\cite{Lin:2017fqe} to address the genuine-fake imbalance. The loss, for a score $s\in\left[0, 1\right]$ and target $t\in \left\{0, 1\right\}$, is:
\begin{equation}
    \mathcal{L}_{\text{focal}} = -
    \begin{cases}
     (1 -\alpha) s^\gamma \log{\left(1 - s\right)} &\text{for } t = 1 \\
     \alpha \left(1 - s\right)^\gamma \log{\left(s\right)} &\text{otherwise}
    \end{cases}\,,
\end{equation}
where the hyperparameter $\gamma$ is set to 2 and $\alpha$ is set to the proportions of fake examples (edges or triplets) in a given graph. The overall edge and triplet loss $\mathcal{L}_{\text{edges}}$ and $\mathcal{L}_{\text{triplets}}$ for the given graph are computed as their respective averages.

\subsection{Track Construction}
\label{pipeline:track_building}

\begin{figure}[!htb]
  \centering
  \includegraphics[width=0.8\linewidth]{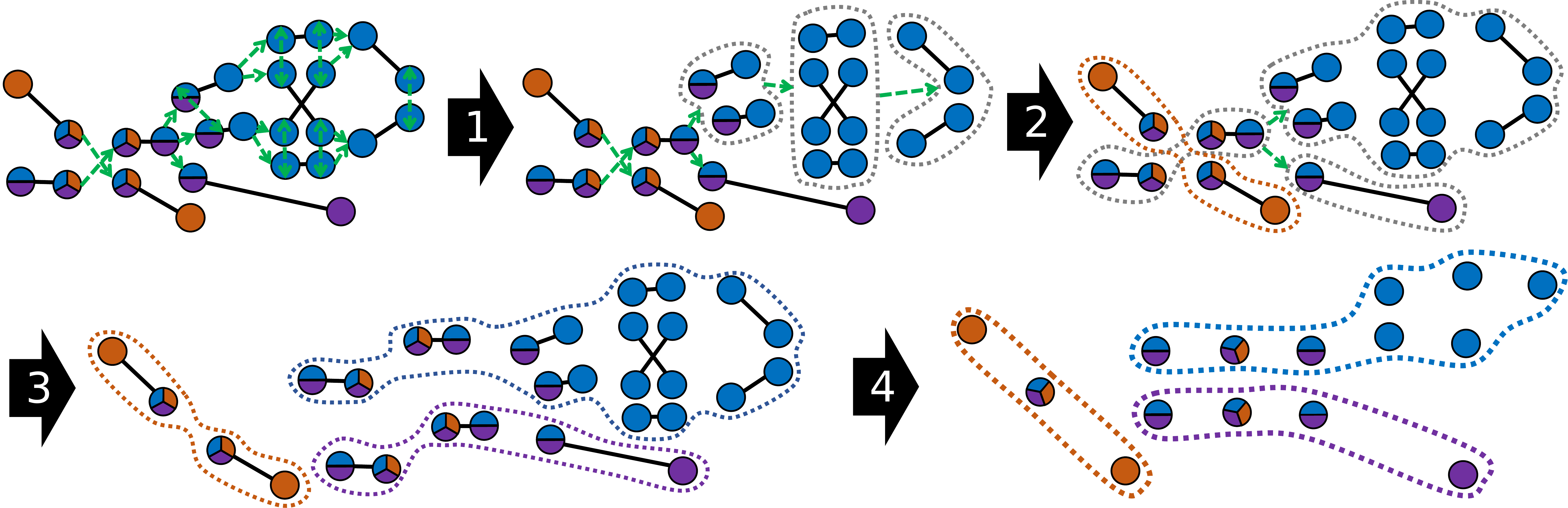}
  \caption{Illustration of the four phases of track construction from the purified edge graph: (1) Connecting left and right elbows, (2) applying a WCC while omitting articulations with shared edges, (3) designating each residual link as a unique track, and (4) substituting edges with hits.}
  \label{fig:track_building}
\end{figure}

Track construction from the purified edge graph $G^{\text{edge}}_{\text{purified}}$ consists of four sequential steps, depicted in Figure~\ref{fig:track_building}:
\begin{enumerate}
    \itemsep0em
    \item Connect left and right elbows. This action eliminates forks (two articulations sharing the same initial or terminal edge) related to a single particle. The residual forks represent two particles that start by sharing hits but later diverge.
    \item Apply a WCC algorithm on the edge graph,  excluding the articulations associated with forks.
    \item Label each remaining link (which corresponds to articulations engaged in forks) as a unique track.
    \item Convert sets of connected edges into corresponding sets of connected hits, representing the tracks.
\end{enumerate}

The initial two steps discern tracks that might intersect. The third step distinguishes tracks that share their first hits but later diverge. By presupposing that tracks sharing multiple successive hits diverge solely once, this method averts sequential iterations within connected components.

\section{Performance}
\label{performance}

\subsection{Definitions and Conventions}
\label{performance:definitions}

The conventions and definitions for track-finding performance in LHCb are outlined in~\cite{Li:2021oga}. A track is matched to a particle when at least 70\% of its hits are associated with that particle, forming a set of \textit{matching candidates} (track, particle). The metrics used to assess the track-finding performance are presented in Table~\ref{tab:metrics}. The uncertainties associated with track-finding efficiency are determined via the Bayesian method with a uniform prior~\cite{Paterno:2004cb} and computed using the \texttt{ROOT} software~\cite{Brun:1997pa}.

\begin{table}[!htb]
\begin{center}
{
\renewcommand{\arraystretch}{1.5}
\begin{tabular}{l|ll}
\hline\hline
Metric & Definition & Formula \\
\hline
Efficiency & Proportion of matched particles & $\frac{\text{\# matched particles}}{\text{\# particles}}$ \\
Clone rate & Proportion of redundant candidates & $\frac{\text{\# candidates - \# matched particles}}{\text{\# candidates}}$ \\
Ghost rate & Proportion of unmatched tracks & $\frac{\text{\# unmatched tracks}}{\text{\# tracks}}$ \\
Hit efficiency & Average proportion of matched hits per particle & $\left\langle\frac{\text{\# matched hits}}{\text{\# hits on particle}}\right\rangle_{\text{candidates}}$ \\
Hit purity & Average proportion of matched hits per track & $\left\langle\frac{\text{\# matched hits}}{\text{\# hits on track}}\right\rangle_{\text{candidates}}$ \\[2mm]
\hline\hline
\end{tabular}
}
\caption{Metrics for track-finding performance. Efficiency, clone rate, and ghost rate encompass all events, while hit efficiency and hit purity are averaged across matching candidates.}
\label{tab:metrics}
\end{center}
\end{table}

Within the Velo's track-finding context, Figure~\ref{fig:tracking_system} illustrates two primary particle categories:
\begin{itemize}
    \itemsep0em
    \item \textbf{Velo-only particles}: particles whose tracks are reconstructible in the Velo but not in the SciFi.
    \item \textbf{Long particles}: particles whose tracks are reconstructible both in the Velo and in the SciFi.
\end{itemize}
Long track trajectories are bent between the UT and the SciFi due to the magnetic field, which enables momentum measurements. Consequently, reconstructing these tracks is crucial for LHCb physics analyses. The aforementioned categories can be further broken down into three sub-categories:
\begin{itemize}
    \itemsep0em
    \item \textbf{No electrons}: All particles except for electrons.
    \item \textbf{Electrons}: Only electrons, which are more challenging to reconstruct due to a higher chance of scattering or photon radiation (bremsstrahlung).
    \item \textbf{From strange}: Non-electron particles in a decay chain with an $s$-quark hadron, excluding electrons. These typically represent tracks originating near the end of the VELO detector, making them harder to reconstruct.
\end{itemize}

\subsection{Training datasets}
\label{performance:datasets}

This study utilises LHCb Run 3 simulation minimum-bias samples of $p$-$p$ collisions, which include overlapping event effects. ``Minimum-bias'' indicates that no specific decay is targeted in the simulation. $p$-$p$ collisions are generated using \texttt{PYTHIA}~\cite{Sjostrand:2007gs}. The interaction of the generated particles with the detector, and its response, are implemented using the \texttt{Geant4} toolkit~\cite{Allison:2006ve} as described in~\cite{Clemencic:2011zza}.

The embedding network and GNN are trained on 700,000 events, adhering to these selection criteria\footnote{Adopting more stringent criteria during the graph construction phase (Section~\ref{pipeline:graph_building}) may reduce the graph's size without compromising performance.}:
\begin{itemize}
    \itemsep0em
    \item The hits of the particle tracks that are not sufficiently linear are removed. This is assessed by fitting a line to the particle hits and applying an upper limit to the average squared distance between the hits and the line. This criterion excludes 2.5\% of the tracks reconstructible in the Velo.
    \item A minimum of 500 genuine Velo hits is required, to offset the removal of these non-linear tracks.
    \item Tracks with fewer than 3 hits are excluded.
\end{itemize}
These criteria are not imposed on the test samples.

\subsection{Parameter Choices}
\label{performance:parameters}

In the graph construction phase, two parameters need to be adjusted: the maximal number of neighbours, $k_{\text{max}}$, and the maximal squared distance in the embedding space, $d^2_{\text{max}}$. $k_{\text{max}}$ is fixed to 50, but it could be reduced for quicker inference.

To select an appropriate value for $d^2_{\text{max}}$, a natural choice might be the margin of the embedding loss $m = 0.010$. To better grasp its impact on track-finding performance, a hit graph is generated for various $d^2_{\text{max}}$ values and then purified by excluding fake edges. Subsequently, a purified edge graph is built, where any fake triplets are discarded. The final tracks emanating from this process provide an upper bound on track-finding performance as a function of \(d^2_{\text{max}}\).

\begin{figure}[!htb]
  \begin{center}
  \subfloat[]{\includegraphics[width=0.47\linewidth]{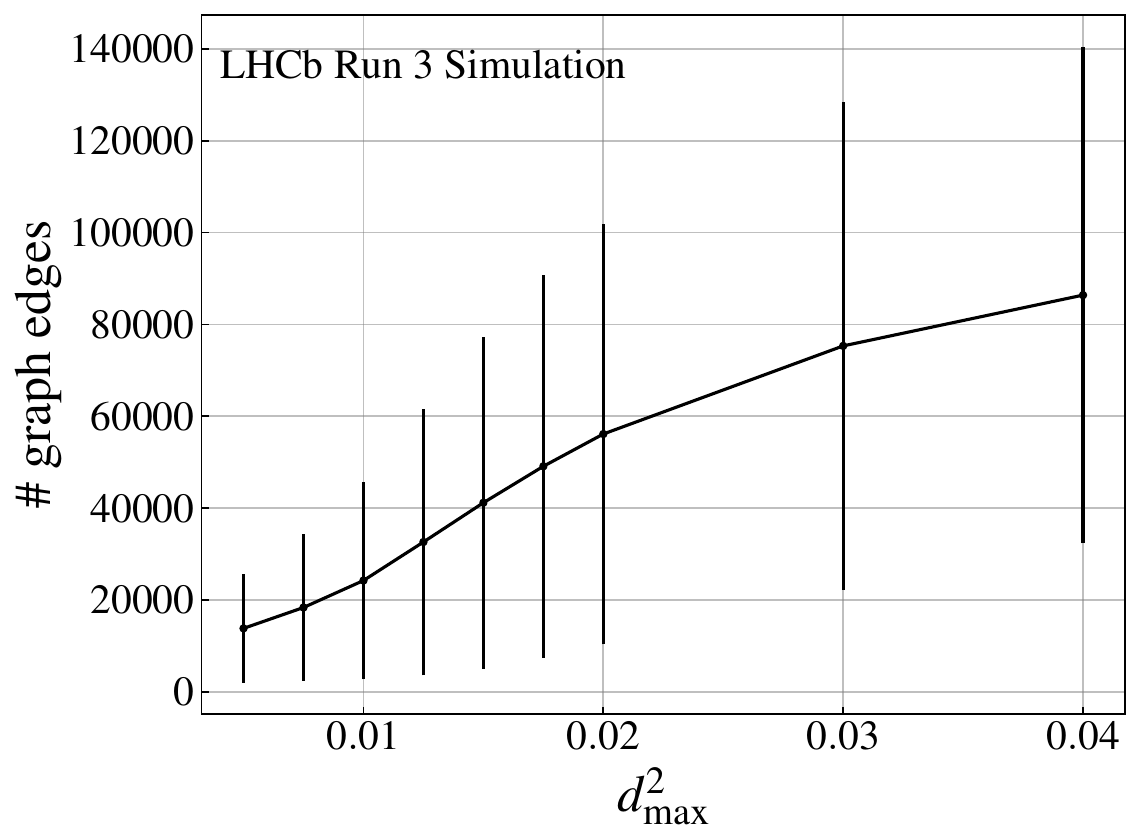}}
  \qquad
  \subfloat[]{\includegraphics[width=0.47\linewidth]{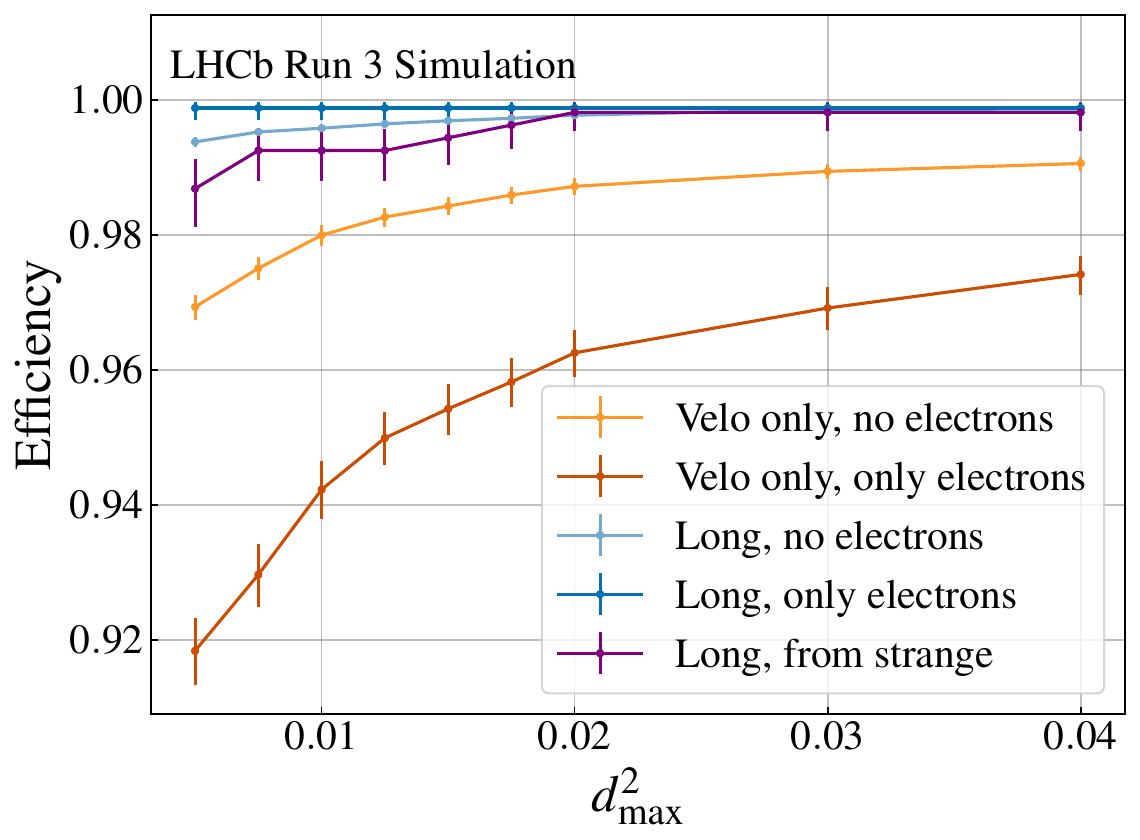}}
  \caption{Comparison of optimal track-finding efficiency (a) and hit graph size (b) based on varying $d^2_\text{max}$, for 200 events. The optimal efficiency, derived after purifying the hit and edge graph, represent the efficiency upper limit according to $d^2_{\text{max}}$. It is shown for 5 different particle categories. Graph sizes (a) feature error bars indicating standard deviation across events, showcasing variability instead of uncertainty.}
  \label{fig:d2max_choice}
  \end{center}
\end{figure}

Figure~\ref{fig:d2max_choice} demonstrates that increasing $d^2_{\text{max}}$ augments graph size (and consequently, inference time), yet can potentially boost performance. Two values are investigated: $d^2_\text{max} = 0.010$ offers a balance between size and performance, while $d^2_\text{max} = 0.020$ elevates efficiency on the challenging long particles from strange decays.

The GNN classifier is trained for each $d^2_{\text{max}}$ value. The selection thresholds, $s_{\text{edge, min}}$ and $s_{\text{triplet, min}}$, used to filter fake edges and triplets, need to be adjusted. While $s_{\text{edge, min}}$ is set to 0.4, it could be fine-tuned to balance throughput and performance. As for $s_{\text{triplet, min}}$, its values are finetuned to ensure a ghost rate below 1\% while maximising efficiency for \textit{long from strange} category. The final thresholds are $s_{\text{triplet, min}} = 0.32$ for $d^2_{\text{max}} = 0.010$ and $s_{\text{triplet, min}} = 0.36$ for $d^2_{\text{max}} = 0.020$.

Performance outcomes based on these parameters are discussed in the subsequent section.

\subsection{Comparison between Allen and \texttt{ETX4VELO}}
\label{performance:comparison}

The search by triplet algorithm in Allen~\cite{CamporaPerez:2021jhc} is compared with \texttt{ETX4VELO} using 5,000 simulated events without any selection. Notably, this comparison does not encompass throughput, which is a critical aspect in Allen.

Table~\ref{tab:long_performance} illustrates that Allen reconstructs long particles well, especially non-electron ones, with an efficiency surpassing 99\%. However, it shows a slight decrease in efficiency for long electrons and for long non-electron particles originating from strange decays. In contrast, with $d^2_\text{max} = 0.010$, \texttt{ETX4VELO} exhibits superior track quality across all categories, indicated by higher hit efficiency and purity. Furthermore, Table~\ref{tab:ghost_rate} demonstrates that the GNN-based pipeline has rate of fake tracks which is over 1\% lower than Allen. In terms of efficiency, \texttt{ETX4VELO} closely matches Allen for the long no electron category, but surpasses it in long electron reconstruction by more than 1.5\%. However, \texttt{ETX4VELO} is slightly behind Allen by 0.2\% for particles from strange decays. When using a larger graph with $d^2_\text{max} = 0.020$, \texttt{ETX4VELO}'s performance improves further, outperforming Allen for long particles from strange decays. 

Similar observations are made for the performance in Velo-only categories, as presented in Table~\ref{tab:velo_only_performance}. Notably, \texttt{ETX4VELO} detects 15\% more Velo-only electrons than the traditional Allen algorithm.

The elevated clone rate for electrons, particularly for long electrons, arises from electron-positron pairs sharing initial hits. The current 70\% matching criterion erroneously matches each track to both particles. Future revisions will refine this criterion for unique track-to-particle matching, thereby better reflecting the algorithm performance.

\begin{table}[!htb]
\begin{center}
\begin{tabular}{l|cccccccccccc}
\hline\hline
Long category & \multicolumn{2}{c}{Efficiency} & \multicolumn{2}{c}{Clone rate} & \multicolumn{2}{c}{Hit efficiency} & \multicolumn{2}{c}{Hit purity} \\
\multicolumn{1}{l|}{} & Allen & \texttt{ETX4VELO} & Allen & \texttt{ETX4VELO} & Allen & \texttt{ETX4VELO} & Allen & \texttt{ETX4VELO} \\
\hline
\multicolumn{1}{l|}{No electrons} & 99.26 & 99.28 (99.51) & 2.54 & 0.96 (0.89) & 96.46 & 98.73 (98.90) & 99.78 & 99.94 (99.94) \\
\multicolumn{1}{l|}{Electrons} & 97.11 & 98.80 (99.22) & 4.25 & 7.42 (7.31) & 95.24 & 96.54 (96.79) & 97.11 & 98.46 (98.46) \\
From strange & 97.69 & 97.50 (98.06) & 2.50 & 0.92 (0.81) & 97.69 & 98.22 (98.77) & 99.34 & 99.68 (99.68) \\
\hline\hline
\end{tabular}
\caption{Track-finding efficiency (in percentages) of the search by triplet algorithm in Allen versus \texttt{ETX4VELO} for long particles. For \texttt{ETX4VELO}, values for both $d^2_\text{max} = 0.010$ (first) and $d^2_\text{max} = 0.020$ (in parentheses) are presented.}
\label{tab:long_performance}
\end{center}
\end{table}

\begin{table}[!htb]
\centering
\begin{tabular}{l|c|cc}
\hline\hline
 & \multirow{2}{*}{Allen} & \multicolumn{2}{c}{ETX4VELO} \\[1mm]
 &  & $d^2_\text{max} = 0.010$ & $d^2_\text{max} = 0.020$ \\[1mm]
 \hline
Ghost rate & 2.18\% & 0.76\% & 0.81\%\\
\hline\hline
\end{tabular}
\caption{Ghost rate of the search by triplet algorithm in Allen versus \texttt{ETX4VELO} for two choices of $d^2_\text{max}$.}
\label{tab:ghost_rate}
\end{table}

\begin{table}[!htb]
\begin{center}
\begin{tabular}{l|cccccccccccc}
\hline\hline
Velo-only & \multicolumn{2}{c}{Efficiency} & \multicolumn{2}{c}{Clone rate} & \multicolumn{2}{c}{Hit efficiency}  & \multicolumn{2}{c}{Hit purity} \\
\multicolumn{1}{l|}{category} & Allen & \texttt{ETX4VELO} & Allen & \texttt{ETX4VELO} & Allen & \texttt{ETX4VELO} & Allen & \texttt{ETX4VELO} \\
\hline
\multicolumn{1}{l|}{No electrons} & 96.84 & 97.03 (97.86) & 3.84 & 1.08 (1.02) & 93.89 & 97.93 (98.32) & 99.50 & 99.84 (99.82) \\
\multicolumn{1}{l|}{Electrons} & 67.81 & 85.10 (86.69) & 10.27 & 5.02 (4.97) & 79.21 & 93.33 (93.88) & 97.35 & 99.07 (98.99) \\
From strange & 93.53 & 93.07 (96.05) & 5.60 & 1.97 (1.77) & 90.05 & 93.92 (96.05) & 99.36 & 99.67 (99.64) \\
\hline\hline
\end{tabular}
\caption{Track-finding efficiency (in percentages) of the search by triplet algorithm in Allen versus \texttt{ETX4VELO} for Velo-only particles. For \texttt{ETX4VELO}, values for both $d^2_\text{max} = 0.010$ (first) and $d^2_\text{max} = 0.020$ (in parentheses) are presented.}
\label{tab:velo_only_performance}
\end{center}
\end{table}

\FloatBarrier
\section{Conclusions}

This work introduced \texttt{ETX4VELO}, a GNN-based pipeline derived from the foundational Exa.TrkX pipeline as presented in~\cite{ExaTrkX:2021abe}. \texttt{ETX4VELO} has the ability to reconstruct tracks sharing hits through an novel triplet-based approach.

When juxtaposed with the default traditional algorithm, \texttt{ETX4VELO} not only matched its efficiency but in some instances, surpassed it. The GNN pipeline notably excels in two areas: it reduces the ghost rate by over 1\% and delivers significantly improved electron reconstruction results. 

As the focus shifts to deployment, the immediate priority is to optimise inference time, to accomodate the high-rate environment that LHCb operates in. Integration of the GNN-based pipeline into Allen is in progress, aiming to provide a comprehensive comparison with the default traditional algorithm. This integration is expected to offer insights that will guide the refinement of various hyperparameters — including the DNN architecture, embedding dimension, number of layers, hidden units, and beyond.

Simultaneously, efforts are directed towards adapting this pipeline for other LHCb tracking detectors, starting from the SciFi.
\newline

The code for training and testing \texttt{ETX4VELO} is accessible at \url{https://gitlab.cern.ch/gdl4hep/etx4velo/-/tree/ctd2023}.


\Acknowledgements
The authors extend their sincere gratitude to the LIP6 laboratory for granting access to their versatile GPU cluster, where the majority of the trainings were conducted. We are grateful to the ANN4Europe group at Frankfurt Institute for Advanced Studies (FIAS), led by Ivan Kisel, for their insightful discussions. We thank LHCb's Real-Time Analysis project for its support, for many useful discussions, and for reviewing an early draft of this manuscript. We also thank the LHCb computing and simulation teams for producing the simulated LHCb samples used to benchmark the performance of the algorithm presented in this paper. The development and maintenance of LHCb's nightly testing and benchmarking infrastructure which our work relied on is a collaborative effort and we are grateful to all LHCb colleagues who contribute to it.
This work is part of the SMARTHEP network and it is funded by the European Union’s Horizon 2020 research and innovation programme, call H2020-MSCA-ITN-2020, under Grant Agreement n. 956086. It is also supported by the ANR-BMBF project ANN4EUROPE ANR-21-FAI1-0011, in collaboration with FIAS.




\begin{thebibliography}{99}


\bibitem{LHCb:2012doh}
    LHCb Collaboration,
    ``Framework TDR for the LHCb Upgrade: Technical Design Report,'' (2012)
    CERN-LHCC-2012-007 URL:\url{https://cds.cern.ch/record/1443882}.

\bibitem{Bediaga:2013tje}
    LHCb Collaboration,
    ``LHCb VELO Upgrade Technical Design Report,'' (2013)
    CERN-LHCC-2013-021 URL:\url{https://cds.cern.ch/record/1624070}.

\bibitem{LHCb:2014uqj}
    LHCb Collaboration,
    ``LHCb Tracker Upgrade Technical Design Report,'' (2014)
    CERN-LHCC-2014-001 URL:\url{https://cds.cern.ch/record/1647400}.

\bibitem{LHCb:2020kay}
    LHCb Collaboration,
    ``LHCb Upgrade GPU High Level Trigger Technical Design Report,'' (2020)
    \href{https://doi.org/10.17181/CERN.QDVA.5PIR}{doi:10.17181/CERN.QDVA.5PIR} URL:\url{https://doi.org/10.17181/CERN.QDVA.5PIR}.

\bibitem{ExaTrkX:2021abe}
    X.~Ju \textit{et al.} [Exa.TrkX],
    ``Performance of a geometric deep learning pipeline for HL-LHC particle tracking,''
    Eur. Phys. J. C \textbf{81} (2021) no.10, 876
    \href{https://doi.org/10.1140/epjc/s10052-021-09675-8}{doi:10.1140/epjc/s10052-021-09675-8}
    [\href{https://arxiv.org/abs/2103.06995}{arXiv:2103.06995 [physics.data-an]}] URL:\url{https://doi.org/10.1140/epjc/s10052-021-09675-8}.

\bibitem{Li:2021oga}
    P.~Li, E.~Rodrigues and S.~Stahl,
    ``Tracking Definitions and Conventions for Run 3 and Beyond,'' (2021)
    LHCb-PUB-2021-005 URL:\url{https://cds.cern.ch/record/2752971}.

\bibitem{Johnson:2021tbd}
    J.~Johnson, M.~Douze, and H.~Jegou,
    ``Billion-Scale Similarity Search with GPUs,''
    IEEE Transactions on Big Data \textbf{7} (2021), 535--547
    \href{https://doi.org/doi:10.1109/TBDATA.2019.2921572}{doi:10.1109/TBDATA.2019.2921572}
    [\href{https://arxiv.org/abs/1702.08734}{arXiv:1702.08734 [cs.CV]}] URL:\url{https://doi.org/doi:10.1109/TBDATA.2019.2921572}.

\bibitem{elfwing_sigmoid-weighted_2018}
    S.~Elfwing, E.~Uchibe, and K.~Doya,
    ``Sigmoid-weighted linear units for neural network function approximation in reinforcement learning,''
    Neural Networks \textbf{107} (2018), 3--11
    \href{https://doi.org/10.1016/j.neunet.2017.12.012}{doi:10.1016/j.neunet.2017.12.012}
    [\href{https://arxiv.org/abs/1702.03118}{ArXiv:1702.03118  [cs.LG]}] URL:\url{https://doi.org/10.1016/j.neunet.2017.12.012}.

\bibitem{ExaTrkX:2020apx}
    N.~Choma \textit{et al.} [Exa.TrkX],
    ``Track Seeding and Labelling with Embedded-space Graph Neural Networks,'' (2020)
    [\href{https://arxiv.org/abs/2007.00149}{arXiv:2007.00149 [physics.ins-det]}] URL:\url{https://arxiv.org/abs/2007.00149}.

\bibitem{Lin:2017fqe}
    T.~Y.~Lin, P.~Goyal, R.~Girshick, K.~He and P.~Doll\'ar,
    ``Focal Loss for Dense Object Detection,'' (2017)
    [\href{https://arxiv.org/abs/1708.02002}{arXiv:1708.02002 [cs.CV]}] URL:\url{https://arxiv.org/abs/1708.02002}.

\bibitem{Paterno:2004cb}
    M.~Paterno,
    ``Calculating efficiencies and their uncertainties,'' (2004)
    \href{https://www.osti.gov/biblio/15017262}{doi:10.2172/15017262} URL:\url{https://lss.fnal.gov/archive/test-tm/2000/fermilab-tm-2286-cd.pdf}.

\bibitem{Brun:1997pa}
    R.~Brun and F.~Rademakers,
    ``ROOT: An object oriented data analysis framework,''
    Nucl. Instrum. Meth. A \textbf{389} (1997), 81-86
    \href{https://doi.org/10.1016/S0168-9002(97)00048-X}{doi:10.1016/S0168-9002(97)00048-X} URL:\url{https://doi.org/10.1016/S0168-9002(97)00048-X}.

\bibitem{Sjostrand:2007gs}
    T.~Sjostrand, S.~Mrenna and P.~Z.~Skands,
    ``A Brief Introduction to PYTHIA 8.1,''
    Comput. Phys. Commun. \textbf{178} (2008), 852-867
    \href{https://doi.org/10.1016/j.cpc.2008.01.036}{doi:10.1016/j.cpc.2008.01.036}
    [\href{https://arxiv.org/abs/0710.3820}{arXiv:0710.3820 [hep-ph]}] URL:\url{https://doi.org/10.1016/j.cpc.2008.01.036}.

\bibitem{Allison:2006ve}
    J.~Allison, K.~Amako, J.~Apostolakis, H.~Araujo, P.~A.~Dubois, M.~Asai, G.~Barrand, R.~Capra, S.~Chauvie and R.~Chytracek, \textit{et al.}
    ``Geant4 developments and applications,''
    IEEE Trans. Nucl. Sci. \textbf{53} (2006), 270
    \href{https://doi.org/10.1109/TNS.2006.869826}{doi:10.1109/TNS.2006.869826} URL:\url{https://doi.org/10.1109/TNS.2006.869826}

\bibitem{Clemencic:2011zza}
    M.~Clemencic \textit{et al.} [LHCb],
    ``The LHCb simulation application, Gauss: Design, evolution and experience,''
    J. Phys. Conf. Ser. \textbf{331} (2011), 032023
    \href{https://doi.org/10.1088/1742-6596/331/3/032023}{doi:10.1088/1742-6596/331/3/032023} URL:\url{https://doi.org/10.1088/1742-6596/331/3/032023}



\bibitem{CamporaPerez:2021jhc}
    D.~H.~C\'ampora P\'erez, N.~Neufeld and A.~Riscos N\'u\~nez,
    ``Search by triplet: An efficient local track reconstruction algorithm for parallel architectures,''
    J. Comput. Sci. \textbf{54} (2021), 101422
    \href{https://doi.org/10.1016/j.jocs.2021.101422}{doi:10.1016/j.jocs.2021.101422}
    [\href{https://arxiv.org/abs/2207.03936}{arXiv:2207.03936 [hep-ex]}] URL:\url{https://doi.org/10.1016/j.jocs.2021.101422}.



\end{thebibliography}
\end{document}